\documentclass[11pt,a4paper]{amsart}
\usepackage[utf8]{inputenc}
\usepackage{amsmath,amssymb, amsthm, amsfonts, gensymb, commath, float, mathtools, enumerate, breqn}
\usepackage{tabularx}
\usepackage{multirow}
\usepackage{esvect}
\usepackage{subcaption}
\usepackage{tikz}
\usetikzlibrary{positioning, arrows, arrows.meta, cd}
\usetikzlibrary{shapes,backgrounds,shapes,positioning,petri,topaths}

\theoremstyle{theorem}
\newtheorem{theorem}{Theorem}[section]

\newtheorem{proposition}[theorem]{Proposition}

\def\beq#1#2\eeq{%
        \begin{equation}%
        \label{#1}%
            #2%
        \end{equation}%
   }

\usepackage{color}
\usepackage{graphicx}
\usepackage{epstopdf}
\usepackage[text={15cm,23cm}]{geometry} 
\usepackage[colorlinks=true,%
            linkcolor=red!50!black,%
            citecolor=blue!50!black,%
            urlcolor=darkgray]{hyperref}  

\theoremstyle{definition}

\title[On spectrum of the harmonic Kronig-Penney model]{On spectrum of the harmonic Kronig-Penney model}

\author{J. Gibbons}
\address{Department of Mathematics, Imperial College London, London SW7 2AZ, UK}
\email{gibbonssoinne@aol.com}

\author{A.P. Veselov}
\address{Loughborough University, Loughborough LE11 3TU, UK}
\email{A.P.Veselov@lboro.ac.uk}

\begin{document}

\maketitle

\medskip

\begin{abstract}
We study the spectral properties of the harmonic modification of the quantum mechanical Kronig-Penney model motivated by the theory of the dressing chain. 
In particular, we show that in the special cases the corresponding ground states can be given explicitly and all spectral gaps have the same length with the ends expressed as zeros of the corresponding Weber-Hermite functions. 
\end{abstract}

\section{Introduction}
The Kronig-Penney model \cite{Alb,K-P} is one of the simplest models of a crystal in quantum mechanics. It is important since it is solvable and describes the electronic band structure similar to those coming from more sophisticated models.

The corresponding quantum Kronig-Penney Hamiltonian is
\beq{KP}
H^{KP}_{a,\gamma}=-\frac{d^2}{dx^2}+2\gamma \sum_{m\in \mathbb Z}\delta(x-(2m+1)a)
\eeq
with Dirac delta-functions located at the lattice points $x\in \Lambda=\{x=(2m+1)a, \,\, m\in \mathbb Z\}.$ 

We will consider the following harmonic modification of the Kronig-Penney model, motivated by the theory of the dressing chain \cite{Shabat, VS}. 
In the most general form the corresponding Schr\"odinger operator has the form 
\beq{hKP}
\mathcal L_{a, \gamma}=-\frac{d^2}{dx^2}+[x^2-1]_{2a}+2\gamma \sum_{m\in \mathbb Z}\delta(x-(2m+1)a),
\eeq
where by $[f(x)]_{2a}$ we denote the $2a$-periodic extension of the function $f(x)$ from the interval $-a< x\leq a$ to the real line.
A special case of $\gamma=\pm a$ will be of particular interest for us.

We derive this special form when tried to find the spectral interpretation of the (discontinuous) periodic solutions $f(x)=[\pm x]_{2a}=f(x+2a)$ of the delay Painlev\'e-I  equation
$$
Tf_x+f_x=Tf^2-f^2+\alpha, \,\, Tf(x)=f(x+2a),
$$
which we studied in \cite{GSV}. Since this equation appeared as a special periodic closure of the dressing chain, the results of \cite{VS} suggested to look whether the corresponding operator has interesting spectral properties. We show that this is indeed the case, proving in particular the following result.

 \begin{theorem}
In the special case $\gamma=\pm a$ the spectrum of the harmonic Kronig-Penney Hamiltonian (\ref{hKP}) has all spectral gaps of the same length $\Delta=2$ with the ends of the gaps given as zeros of the corresponding Weber-Hermite functions. 

The ground states are given explicitly as 
$
\psi_1(x)=[e^{-x^2/2}]_{2a}
$
with $E_1=0$ (for $\gamma=a$) and 
$
\psi_1(x)=[e^{x^2/2}]_{2a}
$
with $E_1=-2$ (for $\gamma=-a$).
\end{theorem}

 Note that for the usual Kronig-Penney Hamiltonian  the length of the spectral gaps $\Delta\sim 2\frac{|\gamma|}{a}$ is constant only asymptotically for large energies (see \cite{Alb} and the next section)
and the energy of the ground state is not explicit, so our harmonic modification demonstrates ``more solvability" than the classical model.
 
 We should mention that a different modification of the Kronig–Penney model with harmonic oscillator wells was recently studied in \cite{MM} by Moore and Marsiglio, who expressed the analytic solutions in terms of the Kummer functions. This paper also provides a good review of the physical literature related to the Kronig-Penney model and its applications in solid state physics.

\section{Kronig-Penney model and its spectrum}

The quantum Hamiltonian of the Kronig-Penney model is given by (\ref{KP}). A rigorous and detailed mathematical treatment of the corresponding spectral problem can be found in Chapter III.2 of \cite{Alb}, which we mainly follow in this section.



The corresponding Bloch-Floquet eigenfunctions satisfy the free equation 
\beq{eqKP}
-\psi''=E \psi, \quad \psi(x+2a)=e^{2i\theta a}\psi(x)
\eeq
outside the lattice points, with the condition on the jump of the derivative on the lattice
\beq{jumpKP}
\psi'(x+)-\psi'(x-)=2\gamma \psi(x), \quad x \in \Lambda.
\eeq
The spectrum is absolutely continuous with the spectral bands consisting of the solutions of the transcendental {\it Kronig-Penney equation} \cite{K-P}
\beq{KPeq}
\cos 2a\sqrt{E}+\frac{\gamma}{\sqrt{E}}\sin 2a\sqrt{E}=\cos 2a\theta.
\eeq

For $\gamma>0$ the spectrum is positive, with the ends of the spectral bands given by $E=k^2$, where $k$ are the roots of the equations 
\beq{endKP}
F_{a,\gamma}(k):=\cos 2ak+\frac{\gamma}{k}\sin 2ak=\pm 1
\eeq
(see the graph below).

\begin{figure} [H]
\includegraphics[scale=0.6]{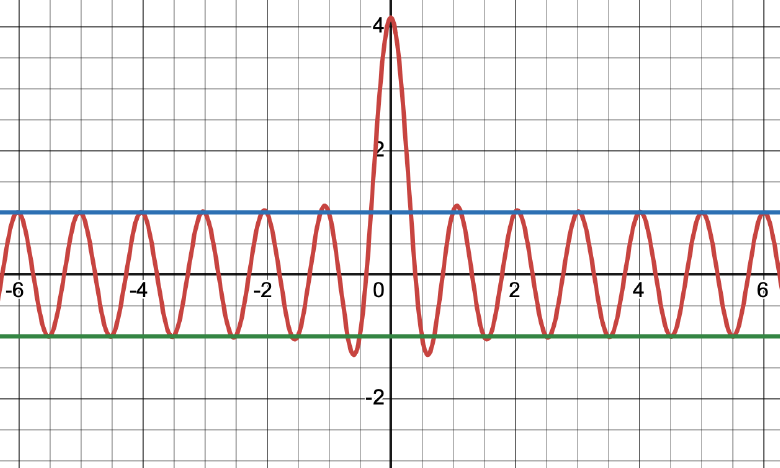} 
\caption{Graph of $F_{a,\gamma}(k)$  with $a=\pi$ and $\gamma=1$.}
\end{figure}


Half of them can be given explicitly as the solutions of $\cos 2ak=\pm 1$: 
$$k=\frac{m\pi}{2a}, \quad E=\frac{m^2\pi^2}{4a^2}, \,\, m\in \mathbb Z_{>0},$$
corresponding to the eigenfunctions $\cos (l+\frac{1}{2})\frac{\pi x}{a}$ and $\sin \frac{l\pi x}{a}$ of the free operator, vanishing at the lattice points.
Indeed, the corresponding equations can be factorised as
$$
1-F_{a,\gamma}(k)=2\sin ak(\sin ak-\frac{\gamma}{k}\cos ak)=0,
$$
$$
1+F_{a,\gamma}(k)=2\cos ak(\cos ak-\frac{\gamma}{k}\sin ak)=0.
$$


The other part consists of $E=k^2,$ where $k$ satisfies one of the transcendental equations
\beq{tank}
\cot ak=\frac{k}{\gamma}, \quad
\tan ak=\frac{k}{\gamma}.
\eeq


For $\gamma<0$ we should consider 3 cases: $a|\gamma|>1$, $a|\gamma|=1$, $a|\gamma|<1$ (see the graphs of the examples with $a=\pi$, $\gamma=-1$ (left) and  $\gamma=-0.25$ (right) below).

\begin{figure} [H]
 \includegraphics[scale=0.4]{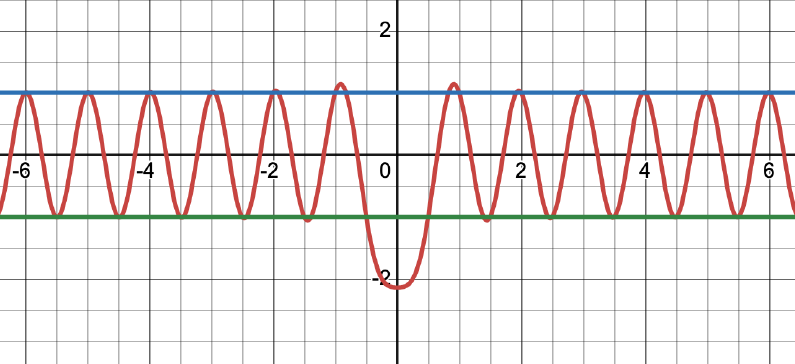}  \quad \includegraphics[scale=0.25]{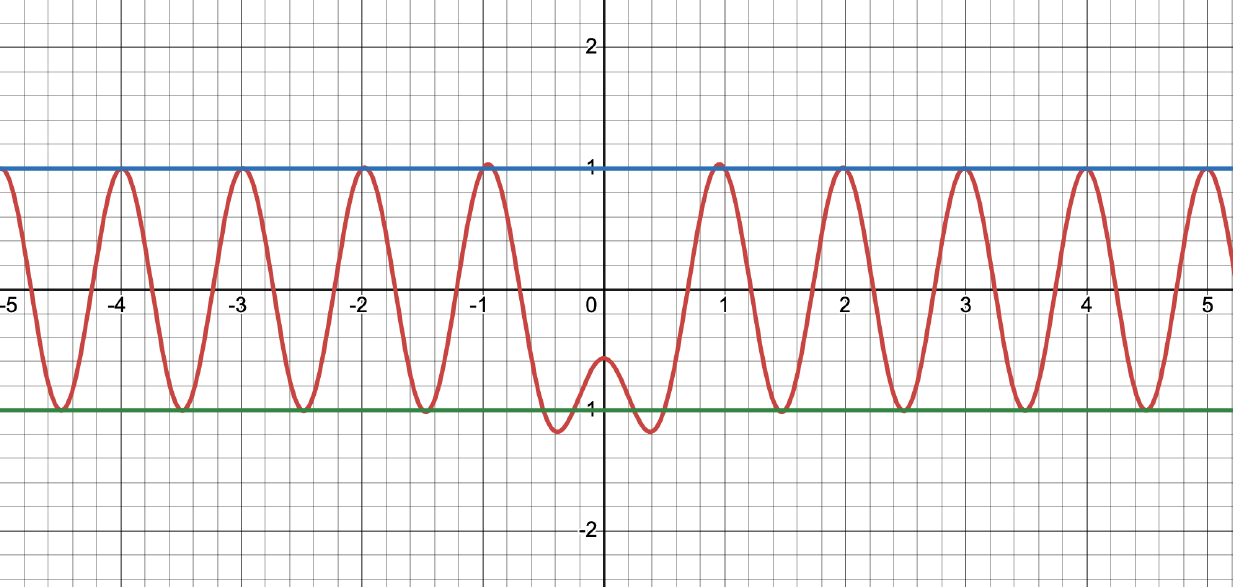} 
\caption{Graphs of $F_{a,\gamma}(k)$  for $\gamma<0$ with $a|\gamma|>1$ (left) and  $a|\gamma|<1$ (right).}
\end{figure}

Let $k_1<k_2<k_3<\dots$ be the ordered joint set of positive roots of $F_{a,\gamma}(k)=\pm 1.$

For negative $\gamma$ we need to consider also the case with $E=-\kappa^2<0$ when the corresponding equations have the form
\beq{endKPneq}
G_{a,\gamma}(k):=\cosh 2ak+\frac{\gamma}{k}\sinh 2ak=\pm 1
\eeq
(see the graphs with $a=\pi$, $\gamma=-0.6$ (left) and  $\gamma=-0.25$ (right) below).

\begin{figure} [H]
 \includegraphics[scale=0.5]{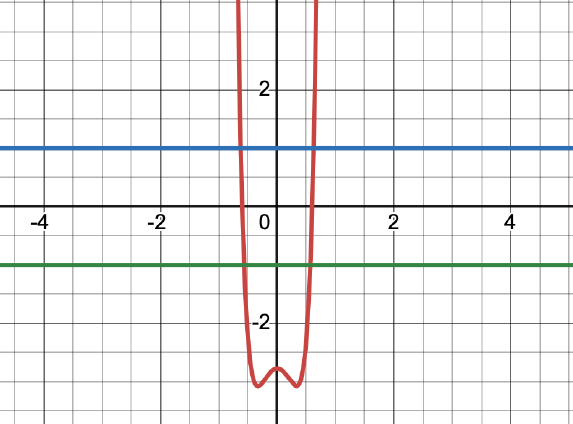}  \quad \includegraphics[scale=0.6]{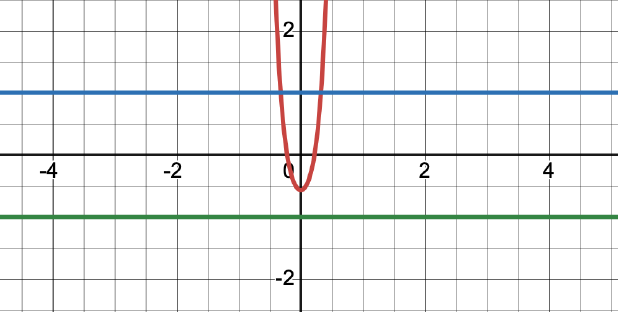} 
\caption{Graphs of $G_{a,\gamma}(\kappa)$  with $\gamma<0$, $a|\gamma|>1$ (left) and  $a|\gamma|<1$ (right).}
\end{figure}

When $a|\gamma|>1$ these equations have two positive roots $0<\kappa_1<\kappa_2$ and 
the spectrum is
\beq{spec2}
\sigma(H^{KP}_{a,\gamma})=\{[-\kappa_2^2,-\kappa_1^2], \cup_{m=1}^\infty \left[k_{2m-1}^2, k_{2m}^2\right ]\}. 
\eeq

If $a|\gamma|<1$ this equation has only one positive root $\kappa_1$ and the spectrum is
\beq{spec22}
\sigma(H^{KP}_{a,\gamma})=\{[-\kappa_1^2,k_1^2], \cup_{m=1}^\infty \left[k_{2m}^2, k_{2m+1}^2\right ]\}. 
\eeq

In the critical case $a|\gamma|=1$ the spectrum is
\beq{spec22}
\sigma(H^{KP}_{a,\gamma})=\{[-\kappa_1^2,0], \cup_{m=1}^\infty \left[k_{2m-1}^2, k_{2m}^2\right ]\}. 
\eeq

The spectrum is strictly monotonic with respect to $\gamma$:
 $$
 \sigma(H^{KP}_{a,\gamma}) \subset \sigma(H^{KP}_{a,\gamma'}), \quad 0<\gamma'<\gamma,
 $$
 $$
 \sigma(H^{KP}_{a,\gamma'})\cap [0,\infty) \subset \sigma(H^{KP}_{a,\gamma})\cap [0,\infty), \quad \gamma'<\gamma<0.
 $$

Note that the ground state $E_1$ is always a solution of a transcendental equation.
 
 The width of the spectral gaps asymptotically is constant:
$
\Delta_m \sim 2|\gamma| a^{-1},
$
while the length of the spectral bands grows linearly with $m$ (see Thm. 2.3.3 in \cite{Alb}).

 \section{Harmonic oscillator and Weber-Hermite functions}
 
 Before considering the harmonic modification of the Kronig-Penney model, let us review the case $\gamma=0, a=\infty,$ when we have the usual harmonic oscillator
 $$
 \mathcal H=-D^2+x^2-1, \quad D=\frac{d^2}{dx^2}.
 $$
  
The corresponding Schr\"odinger equation 
  \beq{Weber}
 -W''+(x^2-1)W=E W
  \eeq
  with general $E$ is known in the classical literature as a particular case of the {\it Weber equation} \cite{WW}. Their solutions $W(x,E)$ are called  the {\it Weber-Hermite (or, parabolic cylinder) functions} and can be expressed in terms of the confluent hypergeometric function \cite{AS}.
  
  The spectrum of the harmonic oscillator is discrete and consists of $E=2n, \, n\in \mathbb Z_{\geq 0}$ with the explicit eigenfunctions
  $$
  W(x,2n)=H_n(x) e^{-x^2/2},
  $$
where $H_n(x)$ are the classical {\it Hermite polynomials}
$
H_n(x)=(-1)^n e^{x^2} D^n e^{-x^2},
$
$$
 H_0=1,\,\,\, H_1=2x,\,\, \, H_2=4x^2-2, \,\, \, H_3=8x^3-12 x, \,\,\, H_4=16x^4-48x^2+12, \dots
$$

The Weber equation has also explicit growing solutions with $E =-2, -4, -6, \dots$:
$$
W(-2n-2,x)=H^*_n(x) e^{x^2/2}, \quad H^*_n(x)=(-i)^n H_n(ix):
$$
$$
 H^*_0=1,\,\,\, H_1^*=2x,\,\, \, H_2^*=4x^2+2, \,\, \, H_3^*=8x^3+12 x, \,\,\, H_4^*=16x^4+48x^2+12, \dots
$$
but all other solutions are non-elementary.

Let $W_{\pm}(x, E)=\pm W(-x,E)$ be the even/odd solutions of the Weber equation (\ref{Weber}).
In the notations of \cite{AS} they are related to the solutions ${\bf y}_1({\bf a},{\bf z}), \, {\bf y}_2({\bf a},{\bf z})$ of the equation
$$
{\bf y}''=({\bf a}+\frac{{\bf z}^2}{4}){\bf y}
$$
by the formulas
\beq{rela}
W_+(x,E)={\bf y}_1(-\frac{E+1}{2},\sqrt 2 x), \quad W_-(x,E)={\bf y}_2(-\frac{E+1}{2},\sqrt 2 x).
\eeq
They can be expressed in terms of the confluent hypergeometric function using 
$$
{\bf y}_1({\bf a},{\bf z})=\exp(-\frac{{\bf z}^2}{4}){}_1 F_1 \left(\frac{1}{2}{\bf a}+\frac{1}{4}; \frac{1}{2}; \frac{{\bf z}^2}{2}\right), \quad {\bf y}_2({\bf a},{\bf z})={\bf z}\exp(-\frac{{\bf z}^2}{4}){}_1 F_1 \left(\frac{1}{2}{\bf a}+\frac{3}{4}; \frac{3}{2}; \frac{{\bf z}^2}{2}\right)
$$
and satisfy the recurrence relations (see Section 19.6 in \cite{AS})
\beq{recur1}
W'_+(x,E)+x W_+(x,E)=-EW_-(x, E-2), \,\, W'_-(x,E)+x W_-(x,E)=W_+(x, E-2),
\eeq
$$
W'_+(x,E)-x W_+(x,E)=-(E+2)W_-(x, E+2), \,\,  W'_-(x,E)-x W_-(x,E)=W_+(x, E+2).
$$

\section{Harmonic modification of the Kronig-Penney model}

Consider now the harmonic Kronig-Penney operator 
\beq{HKP}
\mathcal L_{a, \gamma}=-\frac{d^2}{dx^2}+[x^2-1]_{2a}+2\gamma \sum_{m\in \mathbb Z}\delta(x-(2m+1)a).
\eeq

The solutions of the corresponding Schr\"odinger equation $$\mathcal L_{a, \gamma}\psi(x,E)=E\psi(x,E)$$
must satisfy the Weber equation 
\beq{HKPeq}
-\psi''(x,E)+(x^2-1)\psi(x,E)=E\psi(x,E), \, x \in (-a,a)
\eeq
and the jump conditions  
\beq{jump0}
\psi'(x+,E)-\psi'(x-,E)=2\gamma \psi(x,E)
\eeq
at the lattice points $x\in \Lambda=\{x=(2m+1)a, \,\, m\in \mathbb Z\}.$

By the Bloch-Floquet theory for the periodic Schr\"odinger operators \cite{Alb} the spectrum of $\mathcal L_{a, \gamma}$ is absolutely continuous with the ends of the spectral bands consisting of the eigenvalues $E^{\pm}$ of the spectral problem
$$\mathcal L_{a, \gamma}\psi^{\pm}(x) =E^{\pm}\psi^{\pm}(x)$$
with  periodic and anti-periodic  conditions
$$ \psi^{\pm}(x+2a)= \pm \psi^{\pm}(x).$$

Since the operator has an obvious symmetry $\sigma: x \to -x$ we can consider only even and odd solutions of the corresponding spectral problem.

Let $W_{\pm}(x, E)=\pm W(-x,E)$ be, as before, the even/odd solutions of the Weber equation (\ref{Weber}).
To extend $W_{\pm}(x, E)$ from $x \in [-a,a]$ to a continuous periodic/anti-periodic function on the real line, we need to consider two cases: $W_{\pm}(a, E)=0$ and $W_{\pm}(a, E)\neq 0.$

Let assume first that $W_+(a,E)=0$, then we can extend it to the anti-periodic eigenfunction on the whole line as
$$
\psi^-(x,E)=[W_+(x,E)]_{2a}^{(-)},
$$
where by $[f(x)]_{2a}^{(-)}$ we denote the anti-periodic extension of function $f(x), \, x \in [-a,a]$ with $$f(x+2a)_{2a}^{(-)}=-f(x)_{2a}^{(-)}, \, x \in \mathbb R.$$

Similarly, if $W_-(a,E)=0$ we can extend it to the periodic eigenfunction on the whole line as
$$
\psi^+(x,E)=[W_-(x,E)]_{2a}, \quad \psi^+(x+2a,E)=\psi^+(x,E).
$$
Note that in both cases the jump condition (\ref{jump0})  is satisfied automatically since both sides are zero, and that those are the only possible extensions in this case.

If $W_{\pm}(a, E)\neq 0$ then by continuity condition we have to extend them to the eigenfunctions periodically in the even case: $\psi^{+}(x,E)=[W_{+}(x,E)]_{2a},$ and anti-periodically in the odd case:
 $\psi^{-}(x,E)=[W_{-}(x,E)]^{(-)}_{2a},$
 provided they satisfy the jump condition
\beq{jump1}
W'_\pm(a,E)+\gamma W_\pm(a,E)=0,
\eeq
replacing in this case the Kronig-Penney relation (\ref{endKP}).


Thus we have the following description of the spectrum of the harmonic Kronig-Penney operator, generalising the classical case.

\begin{proposition}
The periodic and anti-periodic ends of the spectral bands of the harmonic Kronig-Penney model are respectively
\beq{perset}
Spec_{per}=\{E: W'_+(a,E)+\gamma W_+(a,E)=0\} \cup \{E: W_-(a,E)=0\} 
\eeq
and 
\beq{aperset}
Spec_{aper}=\{E: W'_-(a,E)+\gamma W_-(a,E)=0\} \cup \{E: W_+(a,E)=0\} 
\eeq
 where $W_\pm(x,E)$ are the corresponding even/odd Weber-Hermite functions.
\end{proposition}

In particular, the ground state of the harmonic Kronig-Penney model is $$\psi_1(x,E_1)=[W_{+}(x,E_1)]_{2a},$$ where $E_1$ is the smallest root of the equation
$
W'_+(a,E)+\gamma W_+(a,E)=0.
$

\section{Special case: $\gamma=\pm a.$}

When $\gamma=a$ we can use the recurrence relations (\ref{recur1}) to write
$$
W'_+(a,E)+\gamma W_+(a,E)=W'_+(a,E)+a W_+(a,E)=-EW_-(a, E-2)
$$
and similarly
$$
W'_-(a,E)+\gamma W_-(a,E)=W_+(a, E-2).
$$

\begin{proposition}
When $\gamma=a$ the ends of the spectral gaps of the harmonic Kronig-Penney model are given by the interlacing roots of
\beq{perset}
W_+(a,E)=0,\,\,  W_+(a,E-2)=0,\,\, W_-(a,E)=0,\,\, W_-(a,E-2)=0.
\eeq
When $\gamma=-a$ the ends of the spectral gaps are given by 
\beq{perset}
W_+(a,E+2)=0,\,\,  W_+(a,E)=0,\,\, W_-(a,E+2)=0,\,\, W_-(a,E)=0.
\eeq

In particular, all spectral gaps have the same length $\Delta=2.$
\end{proposition}

This is in a good agreement with the asymptotic behaviour $\Delta \sim 2|\gamma|/a$ for the usual Kronig-Penney model \cite{Alb}.
Note that in the case of smooth periodic potentials the width of the spectral gaps decays exponentially.

It is interesting also that in contrast to the classical Kronig-Penney model the ground states in these cases can be given explicitly.

\begin{proposition}
In the special case $\gamma=a$ for every $a$ the ground state of the harmonic Kronig-Penney model (\ref{HKP}) can be given explicitly as 
$
\psi_1(x)=[e^{-x^2/2}]_{2a}
$
with $E_1=0$.

Similarly, when $\gamma=-a$ then the ground state is
$
\psi_1(x)=[e^{x^2/2}]_{2a}
$
with $E_1=-2.$
\end{proposition}

Indeed, the corresponding functions $W(x)=e^{\pm x^2/2}$ are positive on the whole line and satisfy the relation $W'=\pm x W.$

\begin{figure} [H]
\includegraphics[scale=0.5]{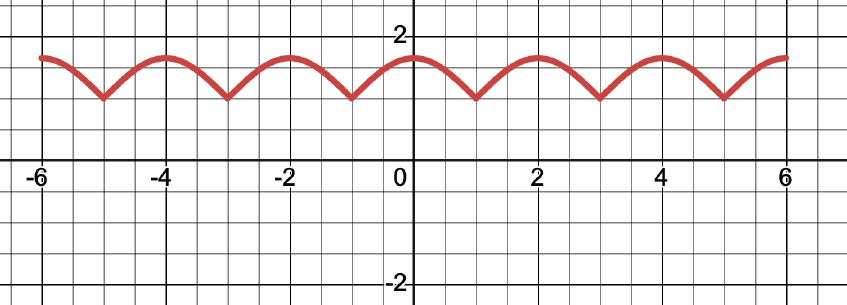} \quad \includegraphics[scale=0.5]{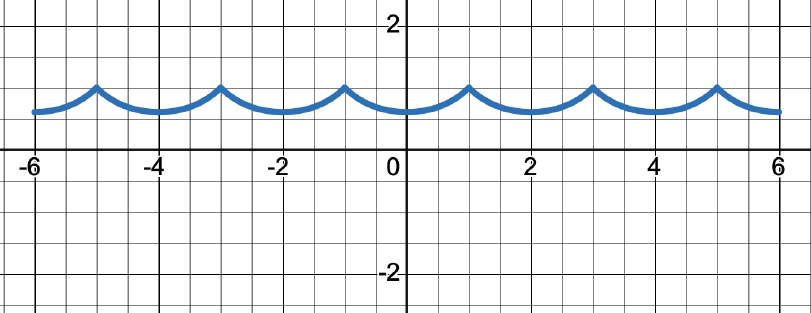} 
\caption{Ground states of the harmonic Kronig-Penney model (\ref{hKP}) with $a=1$ and $\gamma=1$ (left), $\gamma=-1$ (right).}
\end{figure}

If we now specialise $a$ we can describe the first spectral band explicitly.

Let $x_k^*$ be the smallest positive root of the Hermite polynomial $H_k(x).$ Asymptotically for large $k$ we have $x^*_{k} \sim \sqrt{\pi/2k}$.

\begin{proposition}
In the special case $\gamma=a=x^*_{2m}, \, m \in \mathbb N$ the first spectral band of the harmonic Kronig-Penney model  is $$[E_1,E_2]=[0,4m+2]$$ with $\psi_2=[H_{2m+1}(x)e^{-x^2/2}]^{(-)}_{2a}$.

In particular, for $\gamma=a=1/\sqrt{2}$ we have $[E_1,E_2]=[0,6]$ with $\psi_2=[(2x^3-3x)e^{-x^2/2}]^{(-)}_{2a},$
for $\gamma=a=\sqrt{6-2\sqrt 3}/2$ we have $[E_1,E_2]=[0,10]$ with $\psi_2=[(4x^5-20x^3+15x)e^{-x^2/2}]^{(-)}_{2a}.$
\end{proposition}

Indeed, for $W(x)=H_{2m+1}(x)e^{-x^2/2}$ we have
$$
W'(x)=-xW(x)+H_{2m+1}'(x)e^{-x^2/2}=-xW(x)+2(2m+1)H_{2m}(x)e^{-x^2/2},
$$
where we have used the relation $H'_k(x)=2kH_{k-1}(x).$ This means that when $a=\gamma=x^*_{2m}$ we have $W'(a)+\gamma W(a)=0$, and thus the jump condition (\ref{jump1}) is satisfied.
Now from the interlacing property of the zeros of $H_{2m}(x)$  and $H_{2m+1}(x)$ it follows that $H_{2m+1}(x)e^{-x^2/2}$ has its only zero $x=0$ at the interval $[-x^*_{2m}, x^*_{2m}],$ so $[H_{2m+1}(x)e^{-x^2/2}]^{(-)}_{2a}$ is the first excited state with $E=2(2m+1).$

Let now $x^{(m)}_1<x^{(m)}_2<\dots <x^{(m)}_{n}, \, n=[m/2]$ be the ordered set of positive roots of the Hermite polynomial $H_{m}(x).$

\begin{proposition}
In the special case $\gamma=a=x^{(m)}_k$ the harmonic Kronig-Penney model has an explicit level $E=2m+2.$ Here, if $m$ is odd, the corresponding eigenfunction $\psi=[H_{m+1}(x)e^{-x^2/2}]_{2a}$ is periodic with $2k-2$ zeros in the period; if $m$ is even, then $\psi=[H_{m+1}(x)e^{-x^2/2}]^{(-)}_{2a}$  is anti-periodic with $2k-1$ zeros in $(-a,a)$.
\end{proposition}

In particular, for $m=4$ we have 2 positive zeros of $H_4(x)=16x^4-48x^2+12$: $$x^{(4)}_1=\sqrt{6-2\sqrt 3}/2 < x^{(4)}_2=\sqrt{6+2\sqrt 3}/2,$$
and for $a=\sqrt{6+2\sqrt 3}/2$ we have $\psi(x)=[(4x^5-20x^3+15x)e^{-x^2/2}]^{(-)}_{2a}$ with $E=10$, which is one of the ends of the second spectral band.

\section{Explicit levels in other special cases}

Consider now the following special values of $\gamma=\gamma(a),$ 
\beq{gn}
\gamma_{n}:=a-\frac{H'_{n}(a)}{H_{n}(a)}, \quad 0<a<x^*_{n}
\eeq
and
\beq{gn*}
\gamma^*_{n}:=-a-\frac{(H_{n}^*)'(a)}{H_{n}^*(a)}, \quad a>0,
\eeq
where $H_n(x)$ and $H_n^*(x)$ are the corresponding Hermite polynomials, and $x^*_{n}$ as before is the smallest positive root of $H_{n}(x).$
When $n=0$ we have the previous cases $\gamma=\pm a.$

Note that when $n=2k$ is even then $\gamma_{n}(a)>0$ and takes all positive values when $a\in (0,x^*_{n})$, while $\gamma^*_{n}(a)<0$ and takes all negative values for $a>0.$

\begin{proposition}
When $n=2k$ is even and $\gamma=\gamma_{2k}(a)$ the corresponding harmonic Kronig-Penney model has the explicit ground state
$\psi_1=[H_{2k}(x)e^{-x^2/2}]_{2a}$ with energy $E_1=4k.$
Similarly, when $\gamma=\gamma^*_{2k}(a)$ the ground state is
$\psi_1=[H^*_{2k}(x)e^{x^2/2}]_{2a}$ with energy $E_1=-4k-2.$

When $n=2k+1$ is odd and $\gamma=\gamma_{2k+1}(a)$ then
$\psi_2=[H_{2k+1}(x)e^{-x^2/2}]^{(-)}_{2a}$ is the first excited state with $E_2=4k+2.$
Similarly, when $\gamma=\gamma^*_{2k+1}(a)$ the first excited state is
$\psi_2=[H^*_{2k+1}(x)e^{x^2/2}]^{(-)}_{2a}$ with energy $E_2=-4k-4.$

\end{proposition}

The proof follows directly from the jump condition (\ref{jump1}) and properties of the Hermite polynomials. For example, it is known that $H^*_{2k}(x)$ is positive for all $x$, so $W=H^*_{2k}(x)e^{x^2/2}$ is positive, satisfies both the Weber equation with $E=-4k-2$ and the jump condition
$$
W'(a,E)+\gamma W(a,E)=[(H^*)'_{2k}(a)+aH^*_{2k}(a)+\gamma_{2k}^*H^*_{2k}(a)]e^{a^2/2}=0
$$
due to the choice of $\gamma_{2k}^*.$ In the odd case $H^*_{2k+1}(x)$ has its only zero at $x=0$, so we get the first excited state.

In particular, for $n=2$ and $\gamma=\gamma_2=a-\frac{4a}{2a^2-1}, \,\, 2a^2<1$ the ground state is $$\psi_1=[(1-2x^2)e^{-x^2/2}]_{2a}$$ with energy $E_1=4,$
while for $\gamma=-a-\frac{4a}{2a^2+1}$ the ground state is $$\psi_1=[(2x^2+1)e^{x^2/2}]_{2a}$$ with energy $E_1=-6.$

When $n=1$ and $\gamma=\gamma_1=a-a^{-1}$ the first excited state is $$\psi_2=[xe^{-x^2/2}]^{(-)}_{2a}$$ with $E_2=2$, while for $\gamma=\gamma_1^*=-a-a^{-1}$ the first excited state is $$\psi_2=[xe^{x^2/2}]^{(-)}_{2a}$$ with $E_2=-4.$

\section{Concluding remarks}

We have shown that in the special cases the harmonic Kronig-Penney model demonstrates some remarkable spectral properties, not possessed by the usual case.
It would be nice to understand how rare are the singular periodic operators with the spectral gaps of constant length and describe all of them under appropriate analytic assumptions. This can be viewed as a generalisation of the finite-gap theory with zero gap length (see \cite{VS} for the references and its relation with the dressing chain).

The other natural question is about the behaviour in various asymptotic regimes of the general harmonic Kronig-Penney model (\ref{hKP}) with two parameters $a$ and $\gamma$. 

For example, it is natural to expect that for large $a\sim\gamma$ the top of the spectrum with energies $E>>a^2$ behaves like in the usual Kronig-Penney model (in agreement with the classical Plancherel-Rotach asymptotic formula), while the spectral bands at the bottom of the spectrum with the energies $E<<a^2$ shrinking to the discrete spectrum of the harmonic oscillator exponentially fast, but the precise asymptotics requires more investigation. The other interesting regimes include  $a\to \infty$ for fixed $\gamma$  and $\gamma \to \pm \infty$ with fixed $a$.

\end{document}